\documentclass{article} 
\usepackage{iclr2021_conference,times}

\usepackage{amsmath,amsfonts,bm}

\def\eqref#1{equation~\ref{#1}}

\def\1{\bm{1}}

\DeclareMathAlphabet{\mathsfit}{\encodingdefault}{\sfdefault}{m}{sl}
\SetMathAlphabet{\mathsfit}{bold}{\encodingdefault}{\sfdefault}{bx}{n}

\usepackage{hyperref}
\usepackage{url}
\usepackage[dvipdfmx]{graphicx}

\title{Splitting expands the application range of vision transformer \\ Variable Vision Transformer (vViT)}

\author{Takuma Usuzaki, MD \\
Takeda General Hospital\\
3-27 Yamaga-machi Aizuwakamatu\\
Fukushima 965-8585, JAPAN \\
\texttt{takuma.usuzaki.p6@dc.tohoku.ac.jp} \\
}

\iclrfinalcopy 
\begin{document}

\maketitle

\begin{abstract}
Vision Transformer (ViT) has achieved outstanding results in computer vision. Although there are many Transformer-based architectures derived from the original ViT, the dimension of patches are often the same with each other. This disadvantage leads to a limited application range in the medical field because in the medical field, datasets whose dimension is different from each other; e.g. medical image, patients' personal information, laboratory test and so on. To overcome this limitation, we develop a new derived type of ViT termed variable Vision Transformer (vViT). The aim of this study is to introduce vViT and to apply vViT to radiomics using T1 weighted magnetic resonance image (MRI) of glioma. In the prediction of 365 days of survival among glioma patients using radiomics,vViT achieved 0.83, 0.82, 0.81, and 0.76 in sensitivity, specificity, accuracy, and AUC-ROC, respectively. vViT has the potential to handle different types of medical information at once.   
\end{abstract}

\section{Introduction}

Self attention-based architectures, especially Transformer have achieved outstanding results \citep{https://doi.org/10.48550/arxiv.1706.03762}. In this study, Vaswani et al. showed that Transformer reduced the calculation cost while keeping the performance of the recurrent neural network (RNN) in the translation task. Transformer has the potential to supersede other RNN-based models such as long short-term memory and gated RNN \citep{10.1162/neco.1997.9.8.1735,https://doi.org/10.48550/arxiv.1412.3555}. In contrast to the success of Transformer in natural language processing (NLP), convolutional architectures had remained the dominant architecture in computer vision and image recognition \citep{6795724,NIPS2012_c399862d,https://doi.org/10.48550/arxiv.1512.03385}. This situation is partly caused by the difference between NLP and computer vision. In NLP, a token can be easily defined from a sentence. On the other hand, in the application transformer to computer vision, it is a problem with how to define an image that corresponds to the token in NLP. Dosovitskiy et al. gave a solution to this problem by splitting the image into 16×16 images \citep{https://doi.org/10.48550/arxiv.2010.11929}. Dosovitskiy et al. achieved 88.55\% of accuracy using the model termed vision transformer (ViT). Although the performance of ViT was lower than that of some other convolutional architecture-based models, ViT reduces calculation costs and leads to time savings \citep{https://doi.org/10.48550/arxiv.2010.11929}. However, in analyzing a medical image, ViT has a limitation. In medical image analysis, some other features such as patients’ medical information (age, sex, medication history, physical assessment data, laboratory test data, and so on ) and radiomics features as well as the medical image itself. To analyze the information by ViT, inputs and heads dimensions should be controlled because ViT analyzes the splitted images of which dimensions are equal to each other. To overcome this limitation, we constructed a model termed variable vision transformer (vViT) which can handle variable input dimensions. As an example of variable input dimensions, we selected radiomics features regarding each group of features. The radiomics explains quantitative image features from standard-of-care medical imaging\citep{Lambin2017}. The aims of this paper are to introduce vViT and to apply vViT to radiomics using T1 weighted magnetic resonance image (MRI) of glioma.


\section{Methods}

In the model implementation of vViT, we follow the original ViT (Dosovitskiy et al, 2020) and Transformer (Vaswani et al., 2017) as closely as possible. In the model architecture explanation, index $p$ means patch.

\subsection{vViT Model Architecture}
Figure 1 shows an overview of the vViT model architecture. The standard Transformer receives as input a 1D sequence of token embeddings. vViT receives 1D sequence $\bold{x}\in \mathbb{R}$ and split-sequence $\{a_n\} (n\in \mathbb{N}, a_n\in\mathbb{N}, 1\leq n\leq N)$ as an input. The input sequence $\bold{x}$ is divided into $N$ sequences  
$\{\bold{x}_p^1, \bold{x}_p^2, \cdots \bold{x}_p^N\}$ according to split-sequence $\{a_n\}$: the length of $\bold{x}_p^n$ is equal to $a_n (1\leq n\leq N)$. After splitting the input sequence  $\bold{x}$ is divided into $N$ sequences  
$\{\bold{x}_p^1, \bold{x}_p^2, \cdots \bold{x}_p^N\}$, we perform linear transformation by $\bold{D}_n\in R^{a_n\times m}$ to give the same dimension $m$ to all patch.

Similar to  $\texttt{[class]}$  token in Bidirectional Encoder Representations from Transformers (BERT) and ViT, we prepend a learnable embedding to the sequence of embedded patches $(\bold{z}_0^0 = \bold{x}_{\textrm{class}})$, whose output values of the Transformer encoder $(\bold{z}_L^0)$ represents the sequence $\bold{y}$. The classification head is implemented by an MLP with one hidden layer at pre-training time and by a single linear layer at the fine-tuning time based on the previous Transformer study. We prepend the embedded patches $(\bold{z}_0^0 = \bold{x}_{\textrm{class}})$ whose length is $m$.

The Transformer encoder (Vaswani et al., 2017) is composed of alternating layers of multiheaded self-attention (MSA) and multilayer perceptron (MLP) blocks. Layernorm (LN) is applied before every block, and residual connections after every block (Wang et al., 2019; Baevski and Auli, 2019).
We implemented MLP using two layers with a Gaussian Error Linear Unit (GELU) non-linearity.

\begin{align}
\bold{z}_{-1}&=[ 
\bold{x}_p^1\bold{D}_1; \ \bold{x}_p^2\bold{D}_2; \  \cdots; \  \bold{x}_p^N\bold{D}_N]\hspace{10mm}\bold{D}_n\in\mathbb{R}^{a_n\times m} \\
\bold{z}_{0}&=[\bold{x}_{\textrm{class}}; \ 
(\bold{x}_p^1\bold{D}_1)\bold{E}_1; \ (\bold{x}_p^2\bold{D}_2)\bold{E}_2; \  \cdots; \  (\bold{x}_p^N\bold{D}_N)\bold{E}_N]+\bold{E}_{\textrm{pos}} \\
&\hspace{10mm}\bold{E}_n\in\mathbb{R}^{m\times d}, \ \bold{E}_{\textrm{pos}}\in\mathbb{R}^{(N+1)\times d} \nonumber\\
\bold{z}'_{\ell}&=\textrm{MSA}(\textrm{LN}(z_{\ell-1}))+z_{\ell-1}\hspace{10mm} \ell = 1, 2, \cdots, L, \\
\bold{z}_{\ell}&=\textrm{MLP}(\textrm{LN}(z_{\ell}'))+z_{\ell}'\hspace{17.5mm} \ell = 1, 2, \cdots, L, \\
\bold{y}&=\textrm{LN}(z_{L}^0).
\end{align}

\begin{figure}[h]
\begin{center}
\includegraphics[width=150mm]{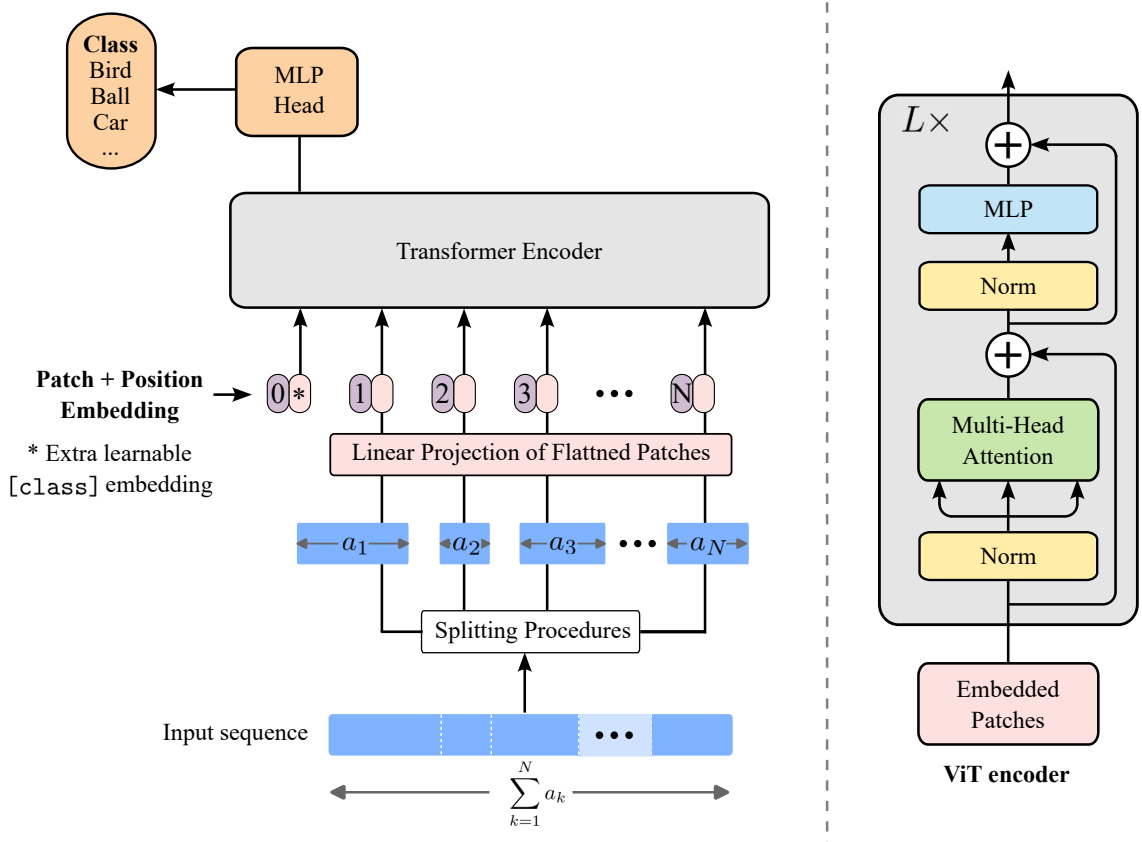}
\end{center}
\caption{vViT model architecture. We split an image into flexible-size patches according to split sequence, linearly embed each of them,
add position embeddings, and feed the resulting sequence of vectors to a standard Transformer
encoder. To perform classification, we add an extra learnable embedding
\texttt{[class]} to the sequence. The illustration of the Transformer encoder was inspired by
\citep{https://doi.org/10.48550/arxiv.1706.03762,https://doi.org/10.48550/arxiv.2010.11929}}
\end{figure}

\subsection{Experiments}

The images used for analysis were obtained from The University of California San Francisco Preoperative Diffuse Glioma MRI (UCSF-PDGM) published in the Image Cancer Archive (TCIA) \citep{doi:10.1148/ryai.220058,Clark2013}. The UCSF-PDGM dataset includes 501 subjects with histopathologically-proven diffuse gliomas who were imaged with a standardized 3 Tesla preoperative brain tumor MRI protocol featuring predominantly 3D imaging, as well as advanced diffusion and perfusion imaging techniques. Multicompartment tumor segmentation of study data was undertaken as part of the 2021 BraTS challenge. Image data first underwent automated segmentation using an ensemble model consisting of prior BraTS challenge-winning segmentation algorithms. Images were then manually corrected by trained radiologists and approved by 2 expert reviewers. Segmentation included three major tumor compartments: enhancing tumor, non-enhancing/necrotic tumor, and surrounding FLAIR abnormality (sometimes referred to as edema). The UCSF-PDGM adds to an existing body of publicly available diffuse glioma MRI datasets that are commonly used in AI research applications. As MRI-based AI research applications continue to grow, new data are needed to foster the development of new techniques and increase the generalizability of existing algorithms. The UCSF-PDGM not only significantly increases the total number of publicly available diffuse glioma MRI cases, but also provides a unique contribution in terms of MRI technique. The inclusion of 3D sequences and advanced MRI techniques like ASL and HARDI provides a new opportunity for researchers to explore the potential utility of cutting-edge clinical diagnostics for AI applications. The details of UCSF-PDGM are explained somewhere else \citep{doi:10.1148/ryai.220058,Clark2013}.

The T1 weighted MRI images and tumor mask images were used to calculate radiomic features. We excluded the images whose mask images contain 16 pixels or lower 16 pixels. After this exclusion, 10,538 images remained. Using PyRadiomics \citep{10.1158/0008-5472.CAN-17-0339}, 104 radiomic features were extraxted; 19 First Order features (Energy, Total Energy, Entropy, Minimum, 10 Percentile, 90 Percentile, Maximum, Mean, Median, Range, Interquartile Range, Mean Absolute Deviation, Robust Mean Absolute Deviation, Root Mean Squared, Standard Deviation, Skewness, Kurtosis, Variance, and Uniformity), 10 Shape2D features (Mesh Surface, Pixel Surface, Perimeter, Perimeter Surface Ratio, Sphericity, Spherical Disproportion, Maximum Diameter, Major Axis Length, Minor Axis Length, and Elongation), 24 Gray-Level Co-occurrence Matrix (GLCM) fetures (Autocorrelation, Joint Average, Cluster Prominence, Cluster Shade, Cluster Tendency, Contrast, Correlation, Difference Average, Difference Entropy, Difference Variance, Joint Energy, Joint Entropy, Informational Measure of Correlation 1, Informational Measure of Correlation 2, Inverse Difference Moment, Maximal Correlation Coefficient, Inverse Difference Moment Normalized, Inverse Difference, Inverse Difference Normalized, Inverse Variance, Maximum Probability, Sum Average, Sum Entropy, and Sum Squares), 16 Gray Level Size Zone Matrix (GLSZM) features (Small Area Emphasis, Large Area Emphasis, Gray Level Non Uniformity, Gray Level Non Uniformity Normalized, Size Zone Non Uniformity, Size Zone Non Uniformity Normalized, Zone Percentage, Gray Level Variance, Zone Variance, Zone Entropy, Low Gray Level Zone Emphasis, High Gray Level Zone Emphasis, Small Area Low Gray Level Emphasis, Small Area High Gray Level Emphasis, Large Area Low Gray Level Emphasis, and Large Area High Gray Level Emphasis), 16 Gray Level Run Length Matrix (GLRLM) features (Short Run Emphasis, Long Run Emphasis, Gray Level Non Uniformity, Gray Level Non Uniformity Normalized, Run Length Non Uniformity, Run Length Non Uniformity Normalized, Run Percentage, Gray Level Variance, Run Entropy, Run Variance, Low Gray Level Run Emphasis, High Gray Level Run Emphasis, Short Run Low Gray Level Emphasis, Short Run High Gray Level Emphasis, Long Run Low Gray Level Emphasis, and Long Run High Gray Level Emphasis), 5 Neighbouring Gray Tone Difference Matrix (NGTDM) features (Busyness, Coarseness, Complexity, Contrast, and Strength), and 14 Gray Level Dependence Matrix (GLDM) features (Small Dependence Emphasis, Large Dependence Emphasis, Gray Level Non Uniformity, Dependence Non Uniformity, Dependence Non Uniformity Normalized, Gray Level Variance, Dependence Variance, Dependence Entropy, Low Gray Level Emphasis, High Gray Level Emphasis, Small Dependence Low Gray Level Emphasis, Small Dependence High Gray Level Emphasis, Large Dependence Low Gray Level Emphasis, and Large Dependence High Gray Level Emphasis). Each group of radiomic features was separately input to vViT.

\subsection{Training and Fine Tuning}

We divided 10,538 slices into the training dataset (7376 slices) and test dataset (3162 slices), and the training dataset was used to train vViT. Binary cross entropy and Adaptive Moment Estimation (Adam) was used for calculating loss and optimizing, respectively. The parameters of Adam were set as learning rate (lr)$=0.0001$, $\beta=(0.9, 0.999)$, $\epsilon=1.0\times 10^{-8}$, weight decay$=0$, and amsgrad=False. The scheduler was used to update the previous lr to $0.99\times$lr at the end of each epoch. The training process was repeated 600 times. After training, the test dataset was used to calculate metrics. 

\subsection{Metrics}

The outcome was defined 1-year survival during follow-up. The information on survival was derived from UCSF-PDGM-metadata.csv published in TCIA. The radiomic feature was evaluated using mean value and standard deviation (SD). Mann–Whitney U test was performed to compare each radiomic feature between the survival and dead group. We calculated sensitivity, specificity, accuracy, and area under the curve of the receiver operating characteristics (AUC-ROC). All analyses were implemented in Python Language (Version 3.8.8).


\section{Results}
Table \ref{train-table1} and \ref{test-table1} in Appendix show the mean, SD, and the result of the Mann–Whitney U test for the training and test, respectively. The best sensitivity, specificity, accuracy, and AUC-ROC were 0.83, 0.82, 0.81, and 0.76, respectively. Figures 2 and 3 show the ROC and changes in sensitivity, specificity, positive predictive value, and negative predicting value, respectively.

\begin{figure}[htbp]
\begin{minipage}{0.45\linewidth}
\vspace*{6mm}
\includegraphics[width=7cm]{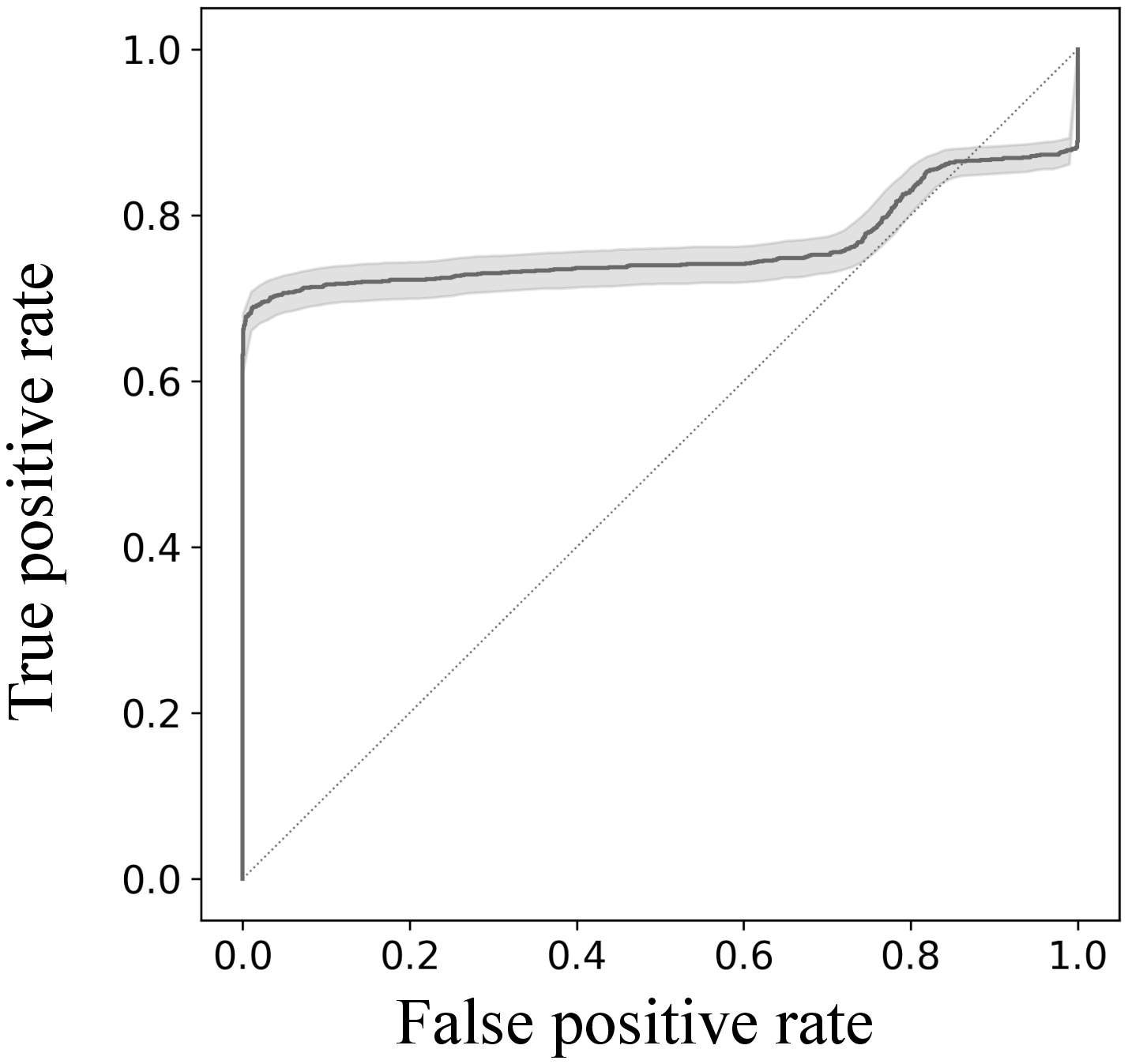}
\caption{ROC is shown. The gray area in the figure shows 95\% confidence interval. The AUC-ROC achieved by vViT was 0.76.}
\end{minipage}
\hspace{1cm}
\begin{minipage}{0.45\linewidth}
\centering
\includegraphics[width=7cm]{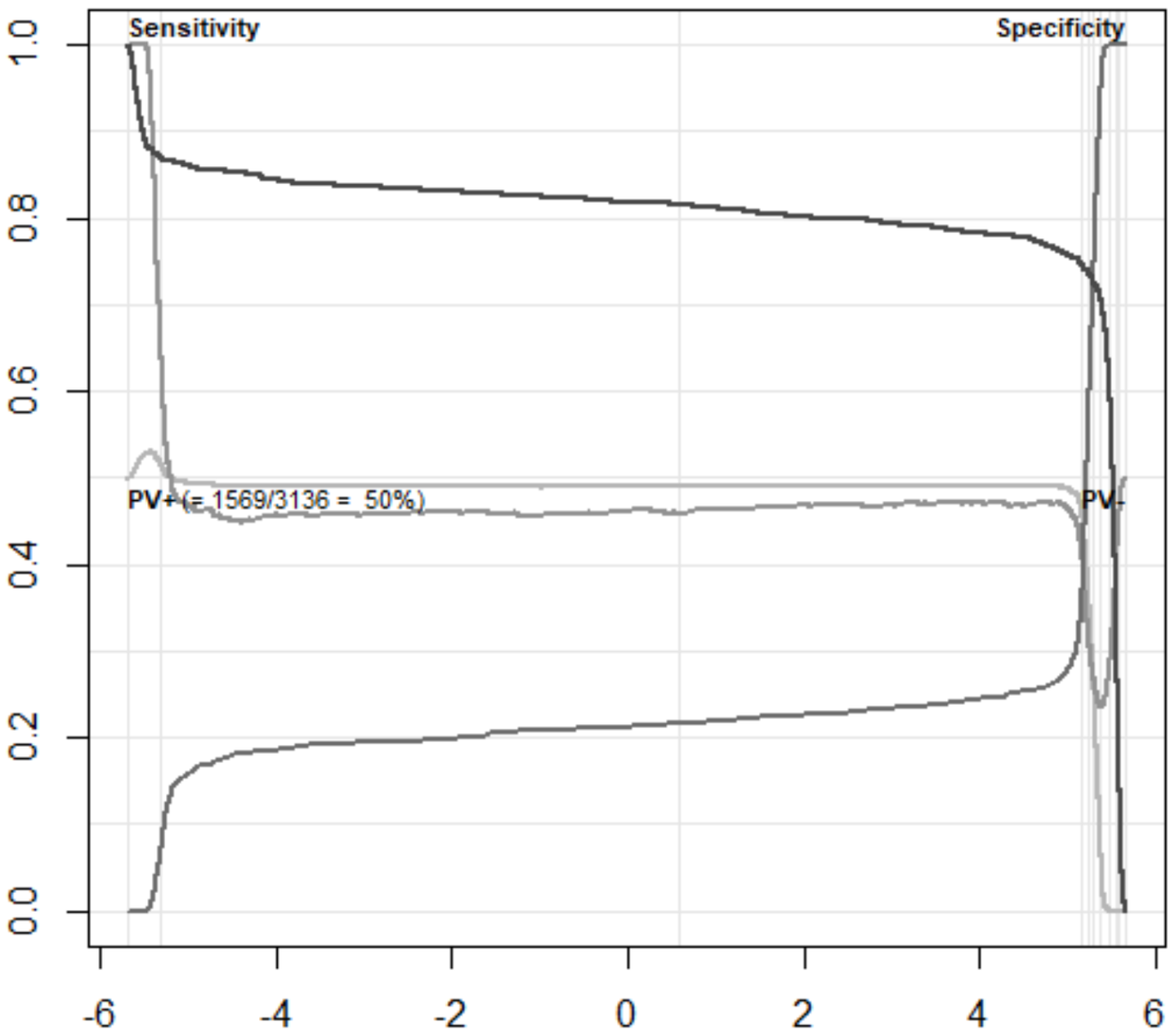}
\caption{The changes in sensitivity, specificity, positive predicting value, and negative predicting value.}
\end{minipage}
\end{figure}

\section{Discussion}

In this paper, we introduced vViT and applied it to predict the 1-year survival of glioma patients using radiomics which was extracted from T1 weighted MRI. vViT achieved 0.81 and 0.76 of accuracy and AUC-ROC, respectively. vViT can handle dataset which contains different dimension subdataset by split-sequence. For example, in this paper, First Order features and Shape 2D features had 19 and 10 features, respectively. When these data are input into a deep learning model, it is controversial how to input into the model. One solution is merging and treating First Order features and Shape 2D features as one sequence. However, this approach leads to losing the information that these features are different from each other. In addition to this, in convolutional architectures, near values in the sequence tend to be convoluted. There is another problem whether these two features are similar to each other. Thus, there is a need for a new architecture that learns features remaining the difference of features. vViT has the potential to be a solution to this problem. 

Some previous studies have examined the association between radiomic features and the prognosis of glioma. Remigio et al. used DeepSurv and PMF-NN models to perform convolutional architectures-based survival analysis in the task of predicting survival at 1050 days \citep{Remigio2022,Katzman2018,https://doi.org/10.48550/arxiv.2111.08239}. DeepSurv and PMF-NN models achieved 0.122 and 0.67 of integrated Brier score and C-index, respectively. Shaheen et al.\citet{10.3389/fnins.2022.911065} examined overall survival using the convolutional architectures-based model. Shaheen et al. reported the best predictive value was 0.73 of AUC-ROC. To our knowledge, in the task of direct prediction of survival among glioma patients, the vViT model achieved state-of-the-art. Other glioma studies have tried to predict mutation such as isocitrate dehydrogenase (IDH) mutation and 1p/19q codeletion using machine learning methods including convolutional architectures \citep{GORE20211599}. The main purpose of these studies is a non-invasive prediction of IDH mutation and 1p/19q codeletion which are associated with the prognosis of glioma. In some of these studies, patients’ medical information was used as well as radiomic features. 

Our study has some limitations. First, patients’ information such as age, sex, medication history, and MRI itself was not included in the input data. Some previous studies revealed the combination of clinical information and the medical image itself contributed to the improvement of deep learning model performance. vViT has the potential to include the information by setting split sequences. Nevertheless, vViT achieved state-of-the-art in the direct prediction of patients’ survival. Further study is needed to solve this limitation. Second, in our analysis, vViT output slice-base prediction and did not output patient base prediction. This means that the prediction may differ depending on each slice. In clinical practice, patient-base prediction is needed. Analyzing the mean probability of final output from vViT may solve this problem. In conclusion, we demonstrated that vViT improved performance in the task of predicting 365 days of survival of glioma patients using radiomics features extracted from T1 weighted MRI. vViT can flexibly handle a dataset that consists of sequences with different dimensions.

\subsubsection*{Author Contributions}
Guarantors of the integrity of the entire study, T.U., K.T.; 
study concepts/study design or data acquisition or data 
analysis/interpretation, T.U., K.T.; manuscript drafting or manuscript 
revision for important intellectual content, T.U., K.T.; approval of final 
version of the submitted manuscript, T.U., K.T.; agree to ensure any 
questions related to the work are appropriately resolved, T.U., K.T.; 
literature research, T.U., K.T.; experimental studies, T.U., K.T.; statistical 
analysis, T.U., K.T.; and manuscript editing, T.U., K.T. T.U.and K.T. equally contributed to this work.

\subsubsection*{Acknowledgments}
The authors acknowledge the National Cancer Institute and the 
Foundation for the National Institutes of Health, and its critical role in 
the creation of the free publicly available UCSF-PDGM database used in this 
study. Data used in this research were obtained from The Cancer Imaging 
Archive (TCIA) sponsored by the Cancer Imaging Program, DCTD/NCI/NIH, 
https://wiki.cancerimagingarchive.net/display/Public/UCSF-PDGM.

\bibliography{iclr2021_conference}

\begin{thebibliography}{16}
\providecommand{\natexlab}[1]{#1}
\providecommand{\url}[1]{\texttt{#1}}
\expandafter\ifx\csname urlstyle\endcsname\relax
  \providecommand{\doi}[1]{doi: #1}\else
  \providecommand{\doi}{doi: \begingroup \urlstyle{rm}\Url}\fi

\bibitem[Adrian et~al.(2022)Adrian, Rick, and Vanessa]{Remigio2022}
Sabariaga~Remigio Adrian, Franich Rick, and Panettieri Vanessa.
\newblock {Application of feed-forward neural network approaches to
  ApplicationApplication of feed-forward neural network approaches to
  radiomics-based survival analysis in glioma patients radiomics-based survival
  analysis in glioma patients}.
\newblock 2022.

\bibitem[Calabrese et~al.(2022)Calabrese, Villanueva-Meyer, Rudie, Rauschecker,
  Baid, Bakas, Cha, Mongan, and Hess]{doi:10.1148/ryai.220058}
Evan Calabrese, Javier~E. Villanueva-Meyer, Jeffrey~D. Rudie, Andreas~M.
  Rauschecker, Ujjwal Baid, Spyridon Bakas, Soonmee Cha, John~T. Mongan, and
  Christopher~P. Hess.
\newblock The university of california san francisco preoperative diffuse
  glioma mri dataset.
\newblock \emph{Radiology: Artificial Intelligence}, 4\penalty0 (6):\penalty0
  e220058, 2022.
\newblock \doi{10.1148/ryai.220058}.
\newblock URL \url{https://doi.org/10.1148/ryai.220058}.
\newblock PMID: 35146430.

\bibitem[Chung et~al.(2014)Chung, Gulcehre, Cho, and
  Bengio]{https://doi.org/10.48550/arxiv.1412.3555}
Junyoung Chung, Caglar Gulcehre, KyungHyun Cho, and Yoshua Bengio.
\newblock Empirical evaluation of gated recurrent neural networks on sequence
  modeling, 2014.
\newblock URL \url{https://arxiv.org/abs/1412.3555}.

\bibitem[Clark et~al.(2013)Clark, Vendt, Smith, Freymann, Kirby, Koppel, Moore,
  Phillips, Maffitt, Pringle, Tarbox, and Prior]{Clark2013}
Kenneth Clark, Bruce Vendt, Kirk Smith, John Freymann, Justin Kirby, Paul
  Koppel, Stephen Moore, Stanley Phillips, David Maffitt, Michael Pringle,
  Lawrence Tarbox, and Fred Prior.
\newblock The cancer imaging archive (tcia): Maintaining and operating a public
  information repository.
\newblock \emph{Journal of Digital Imaging}, 26\penalty0 (6):\penalty0
  1045--1057, Dec 2013.
\newblock ISSN 1618-727X.
\newblock \doi{10.1007/s10278-013-9622-7}.
\newblock URL \url{https://doi.org/10.1007/s10278-013-9622-7}.

\bibitem[Dosovitskiy et~al.(2020)Dosovitskiy, Beyer, Kolesnikov, Weissenborn,
  Zhai, Unterthiner, Dehghani, Minderer, Heigold, Gelly, Uszkoreit, and
  Houlsby]{https://doi.org/10.48550/arxiv.2010.11929}
Alexey Dosovitskiy, Lucas Beyer, Alexander Kolesnikov, Dirk Weissenborn,
  Xiaohua Zhai, Thomas Unterthiner, Mostafa Dehghani, Matthias Minderer, Georg
  Heigold, Sylvain Gelly, Jakob Uszkoreit, and Neil Houlsby.
\newblock An image is worth 16x16 words: Transformers for image recognition at
  scale, 2020.
\newblock URL \url{https://arxiv.org/abs/2010.11929}.

\bibitem[Gore et~al.(2021)Gore, Chougule, Jagtap, Saini, and
  Ingalhalikar]{GORE20211599}
Sonal Gore, Tanay Chougule, Jayant Jagtap, Jitender Saini, and Madhura
  Ingalhalikar.
\newblock A review of radiomics and deep predictive modeling in glioma
  characterization.
\newblock \emph{Academic Radiology}, 28\penalty0 (11):\penalty0 1599--1621,
  2021.
\newblock ISSN 1076-6332.
\newblock \doi{https://doi.org/10.1016/j.acra.2020.06.016}.
\newblock URL
  \url{https://www.sciencedirect.com/science/article/pii/S1076633220303664}.

\bibitem[He et~al.(2015)He, Zhang, Ren, and
  Sun]{https://doi.org/10.48550/arxiv.1512.03385}
Kaiming He, Xiangyu Zhang, Shaoqing Ren, and Jian Sun.
\newblock Deep residual learning for image recognition, 2015.
\newblock URL \url{https://arxiv.org/abs/1512.03385}.

\bibitem[Hochreiter \& Schmidhuber(1997)Hochreiter and
  Schmidhuber]{10.1162/neco.1997.9.8.1735}
Sepp Hochreiter and Jürgen Schmidhuber.
\newblock {Long Short-Term Memory}.
\newblock \emph{Neural Computation}, 9\penalty0 (8):\penalty0 1735--1780, 11
  1997.
\newblock ISSN 0899-7667.
\newblock \doi{10.1162/neco.1997.9.8.1735}.
\newblock URL \url{https://doi.org/10.1162/neco.1997.9.8.1735}.

\bibitem[Katzman et~al.(2018)Katzman, Shaham, Cloninger, Bates, Jiang, and
  Kluger]{Katzman2018}
Jared~L. Katzman, Uri Shaham, Alexander Cloninger, Jonathan Bates, Tingting
  Jiang, and Yuval Kluger.
\newblock Deepsurv: personalized treatment recommender system using a cox
  proportional hazards deep neural network.
\newblock \emph{BMC Medical Research Methodology}, 18\penalty0 (1):\penalty0
  24, Feb 2018.
\newblock ISSN 1471-2288.
\newblock \doi{10.1186/s12874-018-0482-1}.
\newblock URL \url{https://doi.org/10.1186/s12874-018-0482-1}.

\bibitem[Krizhevsky et~al.(2012)Krizhevsky, Sutskever, and
  Hinton]{NIPS2012_c399862d}
Alex Krizhevsky, Ilya Sutskever, and Geoffrey~E Hinton.
\newblock Imagenet classification with deep convolutional neural networks.
\newblock In F.~Pereira, C.J. Burges, L.~Bottou, and K.Q. Weinberger (eds.),
  \emph{Advances in Neural Information Processing Systems}, volume~25. Curran
  Associates, Inc., 2012.

\bibitem[Lambin et~al.(2017)Lambin, Leijenaar, Deist, Peerlings, de~Jong, van
  Timmeren, Sanduleanu, Larue, Even, Jochems, van Wijk, Woodruff, van Soest,
  Lustberg, Roelofs, van Elmpt, Dekker, Mottaghy, Wildberger, and
  Walsh]{Lambin2017}
Philippe Lambin, Ralph~T.H. Leijenaar, Timo~M. Deist, Jurgen Peerlings,
  Evelyn~E.C. de~Jong, Janita van Timmeren, Sebastian Sanduleanu, Ruben~T.H.M.
  Larue, Aniek~J.G. Even, Arthur Jochems, Yvonka van Wijk, Henry Woodruff,
  Johan van Soest, Tim Lustberg, Erik Roelofs, Wouter van Elmpt, Andre Dekker,
  Felix~M. Mottaghy, Joachim~E. Wildberger, and Sean Walsh.
\newblock Radiomics: the bridge between medical imaging and personalized
  medicine.
\newblock \emph{Nature Reviews Clinical Oncology}, 14\penalty0 (12):\penalty0
  749--762, Dec 2017.
\newblock ISSN 1759-4782.
\newblock \doi{10.1038/nrclinonc.2017.141}.
\newblock URL \url{https://doi.org/10.1038/nrclinonc.2017.141}.

\bibitem[LeCun et~al.(1989)LeCun, Boser, Denker, Henderson, Howard, Hubbard,
  and Jackel]{6795724}
Y.~LeCun, B.~Boser, J.~S. Denker, D.~Henderson, R.~E. Howard, W.~Hubbard, and
  L.~D. Jackel.
\newblock Backpropagation applied to handwritten zip code recognition.
\newblock \emph{Neural Computation}, 1\penalty0 (4):\penalty0 541--551, 1989.
\newblock \doi{10.1162/neco.1989.1.4.541}.

\bibitem[Pan et~al.(2021)Pan, Kuo, Rilee, and
  Yu]{https://doi.org/10.48550/arxiv.2111.08239}
Yu~Pan, Kwo-Sen Kuo, Michael~L. Rilee, and Hongfeng Yu.
\newblock Assessing deep neural networks as probability estimators, 2021.
\newblock URL \url{https://arxiv.org/abs/2111.08239}.

\bibitem[Shaheen et~al.(2022)Shaheen, Bukhari, Nadeem, Burigat, Bagci, and
  Mohy-ud Din]{10.3389/fnins.2022.911065}
Asma Shaheen, Syed~Talha Bukhari, Maria Nadeem, Stefano Burigat, Ulas Bagci,
  and Hassan Mohy-ud Din.
\newblock Overall survival prediction of glioma patients with multiregional
  radiomics.
\newblock \emph{Frontiers in Neuroscience}, 16, 2022.
\newblock ISSN 1662-453X.
\newblock \doi{10.3389/fnins.2022.911065}.
\newblock URL
  \url{https://www.frontiersin.org/articles/10.3389/fnins.2022.911065}.

\bibitem[van Griethuysen et~al.(2017)van Griethuysen, Fedorov, Parmar, Hosny,
  Aucoin, Narayan, Beets-Tan, Fillion-Robin, Pieper, and
  Aerts]{10.1158/0008-5472.CAN-17-0339}
Joost~J.M. van Griethuysen, Andriy Fedorov, Chintan Parmar, Ahmed Hosny, Nicole
  Aucoin, Vivek Narayan, Regina~G.H. Beets-Tan, Jean-Christophe Fillion-Robin,
  Steve Pieper, and Hugo~J.W.L. Aerts.
\newblock {Computational Radiomics System to Decode the Radiographic
  Phenotype}.
\newblock \emph{Cancer Research}, 77\penalty0 (21):\penalty0 e104--e107, 10
  2017.
\newblock ISSN 0008-5472.
\newblock \doi{10.1158/0008-5472.CAN-17-0339}.
\newblock URL \url{https://doi.org/10.1158/0008-5472.CAN-17-0339}.

\bibitem[Vaswani et~al.(2017)Vaswani, Shazeer, Parmar, Uszkoreit, Jones, Gomez,
  Kaiser, and Polosukhin]{https://doi.org/10.48550/arxiv.1706.03762}
Ashish Vaswani, Noam Shazeer, Niki Parmar, Jakob Uszkoreit, Llion Jones,
  Aidan~N. Gomez, Lukasz Kaiser, and Illia Polosukhin.
\newblock Attention is all you need, 2017.
\newblock URL \url{https://arxiv.org/abs/1706.03762}.

\end{thebibliography}
\bibliographystyle{iclr2021_conference}

\newpage
\appendix
\section{Appendix}

\vspace*{-0.5cm}
\begin{table}[htbp]
\caption{Radiomic features in the train dataset and results of Mann–Whitney U test}
\label{train-table1}
\begin{center}
\small
\begin{tabular}{llll}
\hline 
Fetures&Survival (SD)&Dead (SD)&p-value\\
\hline
Energy&$6\times 10^9 (7.1\times 10^9)$&$5.5\times 10^9 (6.2\times 10^9)$&0.00024\\
Total Energy&$6\times 10^9 (7.1\times 10^9)$&$5.5\times 10^9 (6.2\times 10^9)$&0.00024\\
Entropy&4.9 (1.1)&4.7 (1.2)&$<$0.0001\\
Minimum&$1.2\times 10^3 (6.3\times 10^2)$&$1\times 10^3 (5.7\times 10^2)$&$<$0.0001\\
10Percentile&$1.6\times 10^3 (6.8\times 10^2)$&$1.4\times 10^3 (6.5\times 10^2)$&$<$0.0001\\
90Percentile&$2.3\times 10^3 (9.6\times 10^2)$&$2\times 10^3 (9.3\times 10^2)$&$<$0.0001\\
Maximum&$2.7\times 10^3 (1.2\times 10^3)$&$2.4\times 10^3 (1.2\times 10^3)$&$<$0.0001\\
Mean&$1.9\times 10^3 (8\times 10^2)$&$1.7\times 10^3 (7.8\times 10^2)$&$<$0.0001\\
Median&$1.9\times 10^3 (8\times 10^2)$&$1.7\times 10^3 (7.8\times 10^2)$&$<$0.0001\\
Range&$1.5\times 10^3 (9.6\times 10^2)$&$1.4\times 10^3 (9.9\times 10^2)$&$<$0.0001\\
Interquartile Range&$3.6\times 10^2 (2.4\times 10^2)$&$3.2\times 10^2 (2.3\times 10^2)$&$<$0.0001\\
Mean Absolute Deviation&$2.1\times 10^2 (1.3\times 10^2)$&$1.9\times 10^2 (1.3\times 10^2)$&$<$0.0001\\
Robust Mean Absolute Deviation&$1.5\times 10^2 (97)$&$1.3\times 10^2 (94)$&$<$0.0001\\
Root Mean Squared&$1.9\times 10^3 (8.1\times 10^2)$&$1.7\times 10^3 (7.9\times 10^2)$&$<$0.0001\\
Standard Deviation&$2.6\times 10^2 (1.6\times 10^2)$&$2.4\times 10^2 (1.6\times 10^2)$&$<$0.0001\\
Skewness&0.11 (0.67)&0.027 (0.77)&$<$0.0001\\
Kurtosis&3.6 (2.7)&3.9 (3.7)&0.014\\
Variance&$9.6\times 10^4 (1.2\times 10^5)$&$8\times 10^4 (8.7\times 10^4)$&$<$0.0001\\
Uniformity&0.054 (0.053)&0.07 (0.071)&$<$0.0001\\
\hline
Mesh Surface&$1.4\times 10^3 (1.1\times 10^3)$&$1.6\times 10^3 (1.2\times 10^3)$&$<$0.0001\\
Pixel Surface&$1.4\times 10^3 (1.1\times 10^3)$&$1.6\times 10^3 (1.2\times 10^3)$&$<$0.0001\\
Perimeter&$2\times 10^2 (1.1\times 10^2)$&$2.2\times 10^2 (1.2\times 10^2)$&$<$0.0001\\
Perimeter Surface Ratio&0.22 (0.18)&0.22 (0.17)&0.038\\
Sphericity&0.67 (0.15)&0.63 (0.15)&$<$0.0001\\
Spherical Disproportion&1.6 (0.42)&1.7 (0.47)&$<$0.0001\\
Maximum Diameter&58 (26)&65 (27)&$<$0.0001\\
Major Axis Length&56 (25)&62 (27)&$<$0.0001\\
Minor Axis Length&34 (16)&36 (17)&$<$0.0001\\
Elongation&0.61 (0.17)&0.6 (0.18)&$<$0.0001\\
\hline
Autocorrelation&$1.4\times 10^3 (1.7\times 10^3)$&$1.3\times 10^3 (1.5\times 10^3)$&$<$0.0001\\
Joint Average&30 (18)&29 (19)&$<$0.0001\\
Cluster Prominence&$3.5\times 10^6 (3.3\times 10^7)$&$1.8\times 10^6 (1.6\times 10^7)$&$<$0.0001\\
Cluster Shade&$8.3\times 10^3 (9.9\times 10^4)$&$3\times 10^3 (3.9\times 10^4)$&$<$0.0001\\
Cluster Tendency&$5.5\times 10^2 (7.5\times 10^2)$&$4.6\times 10^2 (5.2\times 10^2)$&$<$0.0001\\
Contrast&36 (40)&32 (32)&$<$0.0001\\
Correlation&0.82 (0.13)&0.81 (0.13)&$<$0.0001\\
Difference Average&3.8 (2)&3.5 (2.1)&$<$0.0001\\
Difference Entropy&3.2 (0.86)&3 (0.96)&$<$0.0001\\
Difference Variance&17 (23)&15 (16)&$<$0.0001\\
Joint Energy&0.012 (0.026)&0.019 (0.035)&$<$0.0001\\
Joint Entropy&8.1 (1.8)&7.8 (2.1)&$<$0.0001\\
Informational Measure of Correlation 1&-0.34 (0.094)&-0.31 (0.089)&$<$0.0001\\
Informational Measure of Correlation 2&0.96 (0.053)&0.95 (0.065)&$<$0.0001\\
Inverse Difference Moment&0.31 (0.17)&0.35 (0.19)&$<$0.0001\\
Maximal Correlation Coefficient&0.88 (0.075)&0.87 (0.086)&$<$0.0001\\
Inverse Difference Moment Normalized&0.99 (0.01)&0.99 (0.0099)&0.15\\
Inverse Difference&0.39 (0.14)&0.42 (0.16)&$<$0.0001\\
Inverse Difference Normalized&0.94 (0.026)&0.94 (0.025)&0.3\\
Inverse Variance&0.29 (0.12)&0.3 (0.12)&$<$0.0001\\
Maximum Probability&0.029 (0.043)&0.042 (0.064)&$<$0.0001\\
Sum Average&60 (37)&57 (38)&$<$0.0001\\
Sum Entropy&5.7 (1.1)&5.5 (1.3)&$<$0.0001\\
Sum Squares&$1.5\times 10^2 (2\times 10^2)$&$1.2\times 10^2 (1.4\times 10^2)$&$<$0.0001\\
\hline
\end{tabular}
\end{center}
\end{table}

\newpage
\setcounter{table}{0}
\begin{table}[t]
\caption{(continued)}
\label{train-table2}
\begin{center}
\small
\begin{tabular}{llll}
\hline
Fetures&Survival (SD)&Dead (SD)&p-value\\
\hline
Small Area Emphasis&0.74 (0.13)&0.72 (0.14)&$<$0.0001\\
Large Area Emphasis&$49 (1.9\times 10^2)$&$1.5\times 10^2 (8.8\times 10^2)$&$<$0.0001\\
Gray Level Non Uniformity&24 (18)&28 (20)&$<$0.0001\\
Gray Level Non Uniformity Normalized&0.046 (0.041)&0.057 (0.051)&$<$0.0001\\
Size Zone Non Uniformity&$4.7\times 10^2 (4.7\times 10^2)$&$4.9\times 10^2 (5.1\times 10^2)$&0.41\\
Size Zone Non Uniformity Normalized&0.53 (0.16)&0.5 (0.17)&$<$0.0001\\
Zone Percentage&0.61 (0.22)&0.57 (0.24)&$<$0.0001\\
Gray Level Variance&$1.6\times 10^2 (2.1\times 10^2)$&$1.3\times 10^2 (1.4\times 10^2)$&$<$0.0001\\
Zone Variance&$40 (1.7\times 10^2)$&$1.4\times 10^2 (8.5\times 10^2)$&$<$0.0001\\
Zone Entropy&6.2 (0.8)&6.1 (0.83)&$<$0.0001\\
Low Gray Level Zone Emphasis&0.016 (0.034)&0.021 (0.044)&0.031\\
High Gray Level Zone Emphasis&$1.5\times 10^3 (1.7\times 10^3)$&$1.3\times 10^3 (1.5\times 10^3)$&$<$0.0001\\
Small Area Low Gray Level Emphasis&0.011 (0.023)&0.013 (0.025)&0.072\\
Small Area High Gray Level Emphasis&$1.2\times 10^3 (1.5\times 10^3)$&$1.1\times 10^3 (1.3\times 10^3)$&$<$0.0001\\
Large Area Low Gray Level Emphasis&1.6 (12)&$12 (1.1\times 10^2)$&0.00057\\
Large Area High Gray Level Emphasis&$8.9\times 10^3 (2.1\times 10^4)$&$1.1\times 10^4 (2.7\times 10^4)$&$<$0.0001\\
\hline
Short Run Emphasis&0.9 (0.086)&0.88 (0.1)&$<$0.0001\\
Long Run Emphasis&1.7 (1)&2.1 (2)&$<$0.0001\\
Gray Level Non Uniformity&50 (53)&62 (65)&$<$0.0001\\
Gray Level Non Uniformity Normalized&0.052 (0.048)&0.065 (0.062)&$<$0.0001\\
Run Length Non Uniformity&$9.6\times 10^2 (8\times 10^2)$&$1\times 10^3 (8.5\times 10^2)$&0.25\\
Run Length Non Uniformity Normalized&0.78 (0.15)&0.75 (0.17)&$<$0.0001\\
Run Percentage&0.86 (0.11)&0.84 (0.13)&$<$0.0001\\
Gray Level Variance&$1.5\times 10^2 (2\times 10^2)$&$1.3\times 10^2 (1.4\times 10^2)$&$<$0.0001\\
Run Entropy&5.5 (0.79)&5.4 (0.86)&$<$0.0001\\
Run Variance&0.29 (0.44)&0.46 (1)&$<$0.0001\\
Low Gray Level Run Emphasis&0.014 (0.028)&0.019 (0.039)&0.059\\
High Gray Level Run Emphasis&$1.4\times 10^3 (1.7\times 10^3)$&$1.3\times 10^3 (1.5\times 10^3)$&$<$0.0001\\
Short Run Low Gray Level Emphasis&0.012 (0.021)&0.015 (0.028)&0.12\\
Short Run High Gray Level Emphasis&$1.4\times 10^3 (1.6\times 10^3)$&$1.2\times 10^3 (1.4\times 10^3)$&$<$0.0001\\
Long Run Low Gray Level Emphasis&0.033 (0.12)&0.074 (0.3)&0.0086\\
Long Run High Gray Level Emphasis&$1.8\times 10^3 (2\times 10^3)$&$1.7\times 10^3 (1.8\times 10^3)$&0.00012\\
\hline
Busyness&0.17 (0.36)&0.31 (0.69)&$<$0.0001\\
Coarseness&0.028 (0.043)&0.025 (0.038)&$<$0.0001\\
Complexity&$5.1\times 10^3 (9\times 10^3)$&$4.5\times 10^3 (6.3\times 10^3)$&$<$0.0001 \\
Contrast&0.21 (0.39)&0.19 (0.62)&$<$0.0001\\
Strength&29 (44)&22 (33)&$<$0.0001\\
\hline
Small Dependence Emphasis&0.54 (0.2)&0.5 (0.22)&$<$0.0001\\
Large Dependence Emphasis&6.3 (5.8)&7.7 (7.7)&$<$0.0001\\
Gray Level Non Uniformity&68 (85)&$93 (1.2\times 10^2)$&$<$0.0001\\
Dependence Non Uniformity&$4.9\times 10^2 (4.2\times 10^2)$&$5.2\times 10^2 (4.5\times 10^2)$&0.087\\
Dependence Non Uniformity Normalized&0.36 (0.13)&0.35 (0.13)&$<$0.0001\\
Gray Level Variance&$1.5\times 10^2 (1.9\times 10^2)$&$1.3\times 10^2 (1.4\times 10^2)$&$<$0.0001\\
Dependence Variance&1.2 (0.96)&1.4 (1.2)&$<$0.0001\\
Dependence Entropy&6.4 (0.82)&6.3 (0.87)&$<$0.0001\\
Low Gray Level Emphasis&0.014 (0.027)&0.018 (0.038)&0.069\\
High Gray Level Emphasis&$1.4\times 10^3 (1.7\times 10^3)$&$1.3\times 10^3 (1.5\times 10^3)$&$<$0.0001\\
Small Dependence Low Gray Level Emphasis&0.0063 (0.011)&0.0064 (0.011)&0.73\\
Small Dependence High Gray Level Emphasis&$9.6\times 10^2 (1.2\times 10^3)$&$8.7\times 10^2 (1\times 10^3)$&$<$0.0001\\
Large Dependence Low Gray Level Emphasis&0.14 (0.51)&0.28 (0.97)&0.001\\
Large Dependence High Gray Level Emphasis&$4.8\times 10^3 (4.6\times 10^3)$&$4.6\times 10^3 (4.5\times 10^3)$&0.0072\\
\hline
\end{tabular}
\end{center}
\end{table}

\begin{table}[hbtp]
\caption{Radiomic features in the test dataset and results of Mann–Whitney U test}
\label{test-table1}
\begin{center}
\small
\begin{tabular}{llll}
\hline
Fetures&Survival (SD)&Dead (SD)&p-value\\
\hline
Energy&$6.3\times 10^9 (7.5\times 10^9)$&$5.8\times 10^9 (6.4\times 10^9)$&0.0091\\
Total Energy&$6.3\times 10^9 (7.5\times 10^9)$&$5.8\times 10^9 (6.4\times 10^9)$&0.0091\\
Entropy&4.9 (1.1)&4.7 (1.3)&$<$0.0001\\
Minimum&$1.2\times 10^3 (6.2\times 10^2)$&$1\times 10^3 (5.8\times 10^2)$&$<$0.0001\\
10Percentile&$1.6\times 10^3 (6.9\times 10^2)$&$1.4\times 10^3 (6.6\times 10^2)$&$<$0.0001\\
90Percentile&$2.3\times 10^3 (9.6\times 10^2)$&$2\times 10^3 (9.5\times 10^2)$&$<$0.0001\\
Maximum&$2.7\times 10^3 (1.2\times 10^3)$&$2.4\times 10^3 (1.2\times 10^3)$&$<$0.0001\\
Mean&$1.9\times 10^3 (8.1\times 10^2)$&$1.7\times 10^3 (7.9\times 10^2)$&$<$0.0001\\
Median&$1.9\times 10^3 (8.1\times 10^2)$&$1.7\times 10^3 (7.9\times 10^2)$&$<$0.0001\\
Range&$1.6\times 10^3 (9.6\times 10^2)$&$1.4\times 10^3 (9.8\times 10^2)$&$<$0.0001\\
Interquartile Range&$3.6\times 10^2 (2.3\times 10^2)$&$3.2\times 10^2 (2.3\times 10^2)$&$<$0.0001\\
Mean Absolute Deviation&$2.1\times 10^2 (1.3\times 10^2)$&$1.9\times 10^2 (1.3\times 10^2)$&$<$0.0001\\
Robust Mean Absolute Deviation&$1.5\times 10^2 (95)$&$1.3\times 10^2 (94)$&$<$0.0001\\
Root Mean Squared&$1.9\times 10^3 (8.2\times 10^2)$&$1.7\times 10^3 (8\times 10^2)$&$<$0.0001\\
Standard Deviation&$2.7\times 10^2 (1.6\times 10^2)$&$2.4\times 10^2 (1.6\times 10^2)$&$<$0.0001\\
Skewness&0.11 (0.7)&0.033 (0.8)&0.0006\\
Kurtosis&3.6 (2.4)&3.9 (4.2)&0.093\\
Variance&$9.5\times 10^4 (1.1\times 10^5)$&$8.1\times 10^4 (8.8\times 10^4)$&$<$0.0001\\
Uniformity&0.053 (0.051)&0.071 (0.073)&$<$0.0001\\
\hline
Mesh Surface&$1.5\times 10^3 (1.2\times 10^3)$&$1.7\times 10^3 (1.2\times 10^3)$&$<$0.0001\\
Pixel Surface&$1.5\times 10^3 (1.2\times 10^3)$&$1.7\times 10^3 (1.2\times 10^3)$&$<$0.0001\\
Perimeter&$2\times 10^2 (1.1\times 10^2)$&$2.3\times 10^2 (1.2\times 10^2)$&$<$0.0001\\
Perimeter Surface Ratio&0.21 (0.16)&0.22 (0.18)&1\\
Sphericity&0.66 (0.15)&0.63 (0.15)&$<$0.0001\\
Spherical Disproportion&1.6 (0.42)&1.7 (0.46)&$<$0.0001\\
Maximum Diameter&60 (26)&66 (28)&$<$0.0001\\
Major Axis Length&57 (24)&63 (27)&$<$0.0001\\
Minor Axis Length&35 (16)&37 (17)&$<$0.0001\\
Elongation&0.61 (0.16)&0.61 (0.18)&0.62\\
\hline
Autocorrelation&$1.4\times 10^3 (1.7\times 10^3)$&$1.3\times 10^3 (1.4\times 10^3)$&0.00038\\
Joint Average&31 (19)&29 (19)&0.00077\\
Cluster Prominence&$2.8\times 10^6 (2.1\times 10^7)$&$1.9\times 10^6 (1.7\times 10^7)$&$<$0.0001\\
Cluster Shade&$6.2\times 10^3 (6.5\times 10^4)$&$4.2\times 10^3 (4.2\times 10^4)$&0.016\\
Cluster Tendency&$5.5\times 10^2 (6.8\times 10^2)$&$4.6\times 10^2 (5.2\times 10^2)$&$<$0.0001\\
Contrast&37 (36)&33 (32)&0.00024\\
Correlation&0.83 (0.12)&0.81 (0.14)&$<$0.0001\\
Difference Average&3.8 (2.1)&3.5 (2.1)&0.00026\\
Difference Entropy&3.2 (0.86)&3 (0.98)&0.00031\\
Difference Variance&17 (19)&15 (16)&0.00017\\
Joint Energy&0.012 (0.021)&0.02 (0.035)&0.00041\\
Joint Entropy&8.1 (1.8)&7.8 (2.2)&0.00079\\
Informational Measure of Correlation 1&-0.33 (0.09)&-0.31 (0.093)&$<$0.0001\\
Informational Measure of Correlation 2&0.96 (0.05)&0.95 (0.069)&$<$0.0001\\
Inverse Difference Moment&0.31 (0.17)&0.35 (0.2)&$<$0.0001\\
Maximal Correlation Coefﬁcient&0.88 (0.071)&0.87 (0.088)&$<$0.0001\\
Inverse Difference Moment Normalized&0.99 (0.0089)&0.99 (0.01)&0.81\\
Inverse Difference&0.39 (0.14)&0.42 (0.17)&0.00011\\
Inverse Difference Normalized&0.94 (0.024)&0.94 (0.026)&0.73\\
Inverse Variance&0.29 (0.12)&0.3 (0.12)&0.0025\\
Maximum Probability&0.028 (0.041)&0.044 (0.066)&0.0016\\
Sum Average&61 (37)&57 (37)&0.00077\\
Sum Entropy&5.7 (1.1)&5.5 (1.3)&$<$0.0001\\
Sum Squares&$1.5\times 10^2 (1.8\times 10^2)$&$1.2\times 10^2 (1.4\times 10^2)$&$<$0.0001\\
\hline
\end{tabular}
\end{center}
\end{table}

\setcounter{table}{1}
\begin{table}[t]
\caption{(continued)}
\label{test-table2}
\begin{center}
\small
\begin{tabular}{llll}
\hline
Fetures&Survival (SD)&Dead (SD)&p-value\\
\hline
Small Area Emphasis&0.74 (0.13)&0.72 (0.14)&0.0007\\
Large Area Emphasis&$47 (1.7\times 10^2)$&$1.7\times 10^2 (8.9\times 10^2)$&$<$0.0001\\
Gray Level Non Uniformity&25 (19)&29 (20)&$<$0.0001\\
Gray Level Non Uniformity Normalized&0.045 (0.039)&0.058 (0.053)&$<$0.0001\\
Size Zone Non Uniformity&$4.9\times 10^2 (4.8\times 10^2)$&$5.2\times 10^2 (5.2\times 10^2)$&0.51\\
Size Zone Non Uniformity Normalized&0.53 (0.16)&0.5 (0.17)&0.00077\\
Zone Percentage&0.61 (0.22)&0.56 (0.25)&$<$0.0001\\
Gray Level Variance&$1.6\times 10^2 (1.9\times 10^2)$&$1.4\times 10^2 (1.4\times 10^2)$&$<$0.0001\\
Zone Variance&$38 (1.5\times 10^2)$&$1.6\times 10^2 (8.6\times 10^2)$&$<$0.0001\\
Zone Entropy&6.2 (0.76)&6.1 (0.86)&$<$0.0001\\
Low Gray Level Zone Emphasis&0.015 (0.032)&0.021 (0.042)&0.07\\
High Gray Level Zone Emphasis&$1.5\times 10^3 (1.7\times 10^3)$&$1.3\times 10^3 (1.4\times 10^3)$&$<$0.0001\\
Small Area Low Gray Level Emphasis&0.01 (0.019)&0.013 (0.025)&0.11\\
Small Area High Gray Level Emphasis&$1.2\times 10^3 (1.5\times 10^3)$&$1.1\times 10^3 (1.2\times 10^3)$&$<$0.0001\\
Large Area Low Gray Level Emphasis&1.8 (18)&12 (85)&0.0053\\
Large Area High Gray Level Emphasis&$8.2\times 10^3 (1.5\times 10^4)$&$1.2\times 10^4 (2.5\times 10^4)$&0.00056\\
\hline
Short Run Emphasis&0.9 (0.085)&0.87 (0.11)&$<$0.0001\\
Long Run Emphasis&1.7 (0.96)&2.1 (2)&$<$0.0001\\
Gray Level Non Uniformity&52 (56)&67 (67)&$<$0.0001\\
Gray Level Non Uniformity Normalized&0.051 (0.046)&0.067 (0.064)&$<$0.0001\\
Run Length Non Uniformity&$1\times 10^3 (8.2\times 10^2)$&$1.1\times 10^3 (8.7\times 10^2)$&0.2\\
Run Length Non Uniformity Normalized&0.78 (0.15)&0.75 (0.18)&$<$0.0001\\
Run Percentage&0.86 (0.11)&0.84 (0.14)&$<$0.0001\\
Gray Level Variance&$1.5\times 10^2 (1.8\times 10^2)$&$1.3\times 10^2 (1.4\times 10^2)$&$<$0.0001\\
Run Entropy&5.5 (0.77)&5.4 (0.87)&$<$0.0001\\
Run Variance&0.28 (0.41)&0.48 (1)&$<$0.0001\\
Low Gray Level Run Emphasis&0.014 (0.028)&0.019 (0.04)&0.16\\
High Gray Level Run Emphasis&$1.5\times 10^3 (1.7\times 10^3)$&$1.3\times 10^3 (1.4\times 10^3)$&0.00019\\
Short Run Low Gray Level Emphasis&0.011 (0.022)&0.015 (0.028)&0.23\\
Short Run High Gray Level Emphasis&$1.4\times 10^3 (1.6\times 10^3)$&$1.2\times 10^3 (1.4\times 10^3)$&0.00012\\
Long Run Low Gray Level Emphasis&0.034 (0.14)&0.076 (0.31)&0.044\\
Long Run High Gray Level Emphasis&$1.9\times 10^3 (2\times 10^3)$&$1.7\times 10^3 (1.7\times 10^3)$&0.0038\\
\hline
Busyness&0.18 (0.36)&0.35 (0.91)&$<$0.0001\\
Coarseness&0.026 (0.035)&0.024 (0.038)&$<$0.0001\\
Complexity&$5.2\times 10^3 (8.9\times 10^3)$&$4.6\times 10^3 (6.2\times 10^3)$&0.00096\\
Contrast&0.21 (0.38)&0.19 (0.33)&$<$0.0001\\
Strength&29 (38)&22 (28)&$<$0.0001\\
\hline
Small Dependence Emphasis&0.54 (0.2)&0.5 (0.22)&0.00012\\
Large Dependence Emphasis&6.3 (5.7)&8 (8)&$<$0.0001\\
Gray Level Non Uniformity&71 (90)&$1\times 10^2 (1.3\times 10^2)$&$<$0.0001\\
Dependence Non Uniformity&$5.2\times 10^2 (4.4\times 10^2)$&$5.5\times 10^2 (4.6\times 10^2)$&0.087\\
Dependence Non Uniformity Normalized&0.36 (0.13)&0.34 (0.14)&$<$0.0001\\
Gray Level Variance&$1.5\times 10^2 (1.8\times 10^2)$&$1.3\times 10^2 (1.4\times 10^2)$&$<$0.0001\\
Dependence Variance&1.2 (0.95)&1.4 (1.2)&$<$0.0001\\
Dependence Entropy&6.5 (0.78)&6.3 (0.9)&$<$0.0001\\
Low Gray Level Emphasis&0.013 (0.028)&0.019 (0.041)&0.19\\
High Gray Level Emphasis&$1.5\times 10^3 (1.7\times 10^3)$&$1.3\times 10^3 (1.4\times 10^3)$&0.00025\\
Small Dependence Low Gray Level Emphasis&0.0057 (0.0098)&0.0064 (0.012)&0.61\\
Small Dependence High Gray Level Emphasis&$9.9\times 10^2 (1.3\times 10^3)$&$8.7\times 10^2 (9.8\times 10^2)$&$<$0.0001\\
Large Dependence Low Gray Level Emphasis&0.14 (0.56)&0.29 (1.1)&0.015\\
Large Dependence High Gray Level Emphasis&$4.8\times 10^3 (4.5\times 10^3)$&$4.6\times 10^3 (4.4\times 10^3)$&0.058\\
\hline
\end{tabular}
\end{center}
\end{table}
\end{document}